\documentclass[preprint,12pt,nofootinbib]{revtex4}
\usepackage[paper=letterpaper,margin=1.5in]{geometry}
\usepackage{epsfig,color}
\usepackage{graphicx}
\usepackage{amssymb,amsmath}
\usepackage{time}
\usepackage[varg]{txfonts}
\usepackage{hyperref}
\hypersetup{
  colorlinks,
  citecolor=green,
  linkcolor=blue}
\newcommand{\be}{\begin{equation}}
\newcommand{\ee}{\end{equation}}
\newcommand{\bc}{\begin{center}}
\newcommand{\ec}{\end{center}}

\renewcommand{\vec}[1]{\textnormal{\boldmath$#1$}}
\begin{document}

\bibliographystyle{revtex}

\begin{flushright}
{\normalsize
DESY-15-134\\
SLAC-PUB-16353\\
August 2015}
\end{flushright}

\vspace{.4cm}

\title
{Calculation of Wakefields in 2D Rectangular Structures\footnote{This material 
is based upon work supported by the U.S. Department of Energy, Office of 
Science, Office of Basic Energy Sciences, under Contract No. DE-AC02-76SF00515.}
}
\author{I. Zagorodnov}
\affiliation{Deutsches Elektronen-Synchrotron, Notkestrasse 85,
22603 Hamburg, Germany}
\author{K.L.F.~Bane, G. Stupakov}
\affiliation{SLAC National Accelerator Laboratory, Stanford University, 
Stanford, CA 94309\vspace{.4cm}}

\vspace{.4cm} 
\begin{abstract} 
We consider the calculation of electromagnetic fields generated by an electron bunch passing through a vacuum chamber structure that, in general, consists of an entry pipe, followed by some kind of transition or cavity, and ending in an exit pipe. We limit our study to structures having rectangular cross-section, where the height can vary as function of longitudinal coordinate but the width and side walls remain fixed. For such structures, we derive a Fourier representation of the wake potentials through one-dimensional functions. 
A new numerical approach for calculating the wakes in such structures is proposed and implemented in the computer code ECHO(2D).
The computation resource requirements for this approach are moderate and 
comparable to those for finding the wakes in 2D rotationally symmetric 
structures. Numerical examples obtained with the new numerical code are 
presented.

\vfill \centerline
{Submitted to Physical Review Special Topics--Accelerators and Beams}
\end{abstract}

\maketitle


\section{Introduction}

The interaction of charged particle beams and the vacuum chamber environment can be quantified using the concept of impedance or wakefield~\cite{Zotter98}.	
In order to calculate electromagnetic fields in accelerators several different numerical approaches have been suggested and implemented in computer codes. Such calculations for complicated three dimensional structures, however, remain a challenge even for today's parallel computers. The approaches suggested in papers~\cite{Pukhov99}-\cite{Zag05} allow one to obtain reliable results for rotationally symmetric and general three-dimensional structures even when using a personal computer. However, fully three dimensional calculations can be cumbersome and require long computation times. In contrast, for rotationally symmetric structures, the modelling is much simpler, the execution time is much shorter, and the required computational resources are relatively modest.

\begin{figure}[htbp]
\centering
\includegraphics*[height=120mm]{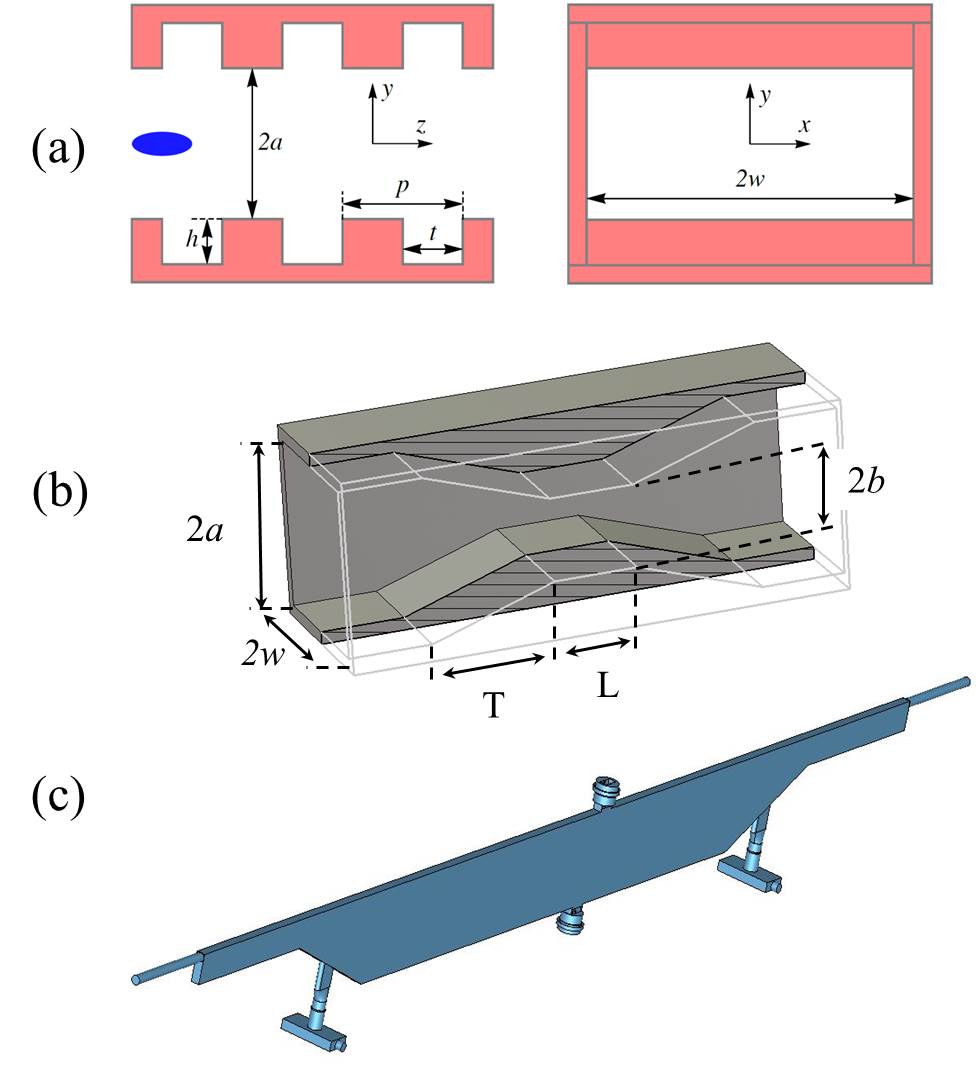}
\caption{Structures of constant width: (a) dechirper, (b) tapered collimator, (c) vacuum chamber in bunch compressor.}\label{Fig01}
\end{figure}

Many three-dimensional vacuum chamber components used in particle accelerators can be well-approximated with structures of rectangular cross-section whose height can vary as function of longitudinal coordinate but whose width and side walls remain fixed. Three examples of such geometry are shown in Fig.~\ref{Fig01}:  (a) a corrugated structure that can be used as a dechirper ~\cite{Bane12}, (b) a rectangular tapered collimator,  and (c) a vacuum chamber in a bunch compressor. Calculation of the wakefields for such structures can be greatly simplified by expanding the field generated by the beam into Fourier series. As it turns out, each Fourier harmonic can be solved separately, and the wakefield calculated as a sum of their contributions. This reduces the original 3D problem into a number of 2D ones, each of which requires much less computing resources than the original problem.

In this paper we derive the equations for the Fourier harmonics and propose a computational scheme for their solution. The scheme was implemented in a computer code that we call ECHO(2D).

The paper is organized as follows. In Section~\ref{sec:2}, we apply the Fourier transform in the horizontal direction (the direction of constant width), and derive a system of equations for the Fourier harmonics. In Section~\ref{sec:3}, we prove that the longitudinal wake function satisfies the Laplace equation with respect to the coordinates of the source particle. In Section~\ref{sec:4}, we show that, in the general case, the wakefield of each harmonic can be described by four scalar functions of one variable. The description is further simplified in Section~\ref{sec:5} where we assume that the entire structure has a horizontal symmetry plane, and then analyze the longitudinal and transverse wakefields near the system axis. An important symmetry relation for the wake function is established in Section~\ref{sec:6}. In Sections~\ref{sec:7} and~\ref{sec:8} we present a numerical algorithm for solving equations derived in Section~\ref{sec:2}. Several numerical examples obtained with the computer code ECHO(2D) are described in Section~\ref{sec:9}. The results of the paper are summarized in Section~\ref{sec:10}.

%
\section{Problem formulation and Fourier expansion}\label{sec:2}
%

\begin{figure}[htbp]
\centering
\includegraphics*[height=40mm]{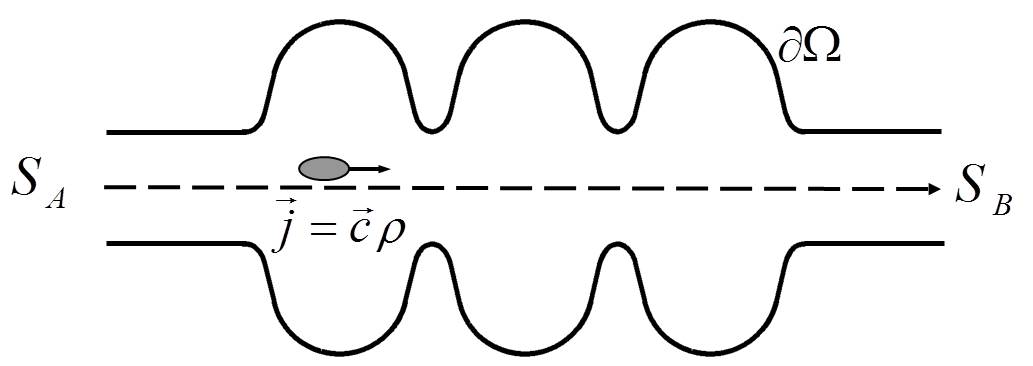}
\caption{Charged particle bunch moving through an accelerating structure.}\label{Fig02}
\end{figure}

We consider a bunch of particles moving with the velocity of light $c$ through a metallic structure comprising an incoming pipe, a transition or cavity, and an outgoing pipe, an example of which is shown in Fig.~\ref{Fig02}. The bunch is characterized by the charge distribution  $\rho$ and the electric current density $\vec{j}=\vec{c}\rho$. We assume that the bunch is moving along a straight line parallel to the longitudinal axis of the system, and we neglect the influence of the wakefields on the bunch's motion. Our problem is to  find the electric and magnetic fields $\vec{E}$, $\vec{H}$, in the domain $\Omega$   which is bounded transversally by perfectly conducting metallic walls, defined by the boundary $\partial\Omega$ of the domain $\Omega$. We need to solve Maxwell's equations with the boundary conditions:
\begin{align}\label{Eq01}
  \nabla\times \vec{H}&=\frac{\partial}{\partial t}\vec{D}+\vec{j},
  &\nabla\times \vec{E}&=-\frac{\partial}{\partial t}\vec{B},\nonumber\\
  \nabla \cdot \vec{D}&=\rho,
  &\nabla \cdot \vec{B}&=0,\nonumber\\
  \vec{B}& = \mu\vec{H},
  &\vec{D} &= \epsilon\vec{E}, \qquad\vec{r}\in\Omega,\nonumber\\
  \vec{E}(t=0)&=\vec{E_0},
  &\vec{H}(t=0)&=\vec{H_0}, \qquad\vec{r}\in\bar\Omega,\nonumber\\
  \vec{n}\times\vec{E}&=0, \qquad x \in \partial\Omega,
\end{align}
where $\vec{E_0}$, $\vec{H_0}$, is an initial electromagnetic field in the domain $\bar\Omega$ and $\vec{n}$ is a unit vector normal to the surface $\partial\Omega$. While we consider the propagation of beams through vacuum, for the generality of the computational algorithm we included in Eqs.~\eqref{Eq01} arbitrary dielectric constant $\epsilon$ and magnetic permittivity $\mu$.


We choose a coordinate system with $y$ in the vertical and $x$ in the horizontal directions; the $z$ coordinate is directed along the longitudinal axis of the system. The structures considered in this paper have constant width $2w$ and a fixed location of the side walls (in the $x$-direction). 

The charge density is localized within the interval $0<x<2w$ and vanishes at the ends of the interval, $\rho(0,y,z)=\rho(2w,y,z)=0$. Such a function can be expanded in Fourier series
\begin{align}\label{Eq03}
	\rho(x,y,z)&=\frac{1}{w}\sum\limits_{m=1}^\infty \rho_m(y,z)\sin(k_{x,m} x),\qquad k_{x,m} =\frac{\pi}{2w}m,\nonumber\\
	\rho_m(y,z)&=\int_{0}^{2w}\rho(x,y,z)\sin(k_{x,m} x)dx.
\end{align}
It follows from the linearity of Maxwell's equations, Eqs.~(\ref{Eq01}) and Eq.~(\ref{Eq03}), that the components of electromagnetic field can be represented by infinite sums:
\begin{align}\label{Eq04}
  \begin{pmatrix}
	H_x\\E_y\\E_z
	\end{pmatrix}
	=\frac{1}{w}\sum\limits_{m=1}^\infty
	\begin{pmatrix}
	H_{x,m}\\E_{y,m}\\E_{z,m}
	\end{pmatrix}
	\sin(k_{x,m} x),\qquad
	\begin{pmatrix}
	E_x\\H_y\\H_z
	\end{pmatrix}
	=\frac{1}{w}\sum\limits_{m=1}^\infty
	\begin{pmatrix}
	E_{x,m}\\H_{y,m}\\H_{z,m}
	\end{pmatrix}
	\cos(k_{x,m} x).
\end{align}
For each mode number $m$   we can write an independent system of equations
\begin{align}\label{Eq05}
 &\frac{\partial}{\partial y} H_{z,m}-\frac{\partial}{\partial z}H_{y,m}=j_{x,m}+\frac{\partial}{\partial t}E_{x,m}\epsilon,\nonumber\\
 &\frac{\partial}{\partial z}H_{x,m} +k_{x,m}H_{z,m}=j_{y,m}+\frac{\partial}{\partial t}E_{y,m}\epsilon,\nonumber\\
 &-k_{x,m}H_{y,m}-\frac{\partial}{\partial y} H_{x,m}=j_{z,m}+\frac{\partial}{\partial t}E_{z,m}\epsilon,\nonumber\\
 &\frac{\partial}{\partial y} E_{z,m}-\frac{\partial}{\partial z}E_{y,m}=-\frac{\partial}{\partial t}H_{x,m}\mu,\nonumber\\
 &\frac{\partial}{\partial z}E_{x,m}-k_{x,m}E_{z,m}=-\frac{\partial}{\partial t}H_{y,m}\mu,\nonumber\\
 &k_{x,m}E_{y,m}-\frac{\partial}{\partial y} E_{x,m}=-\frac{\partial}{\partial t}H_{z,m}\mu,\nonumber\\
 &k_{x,m}H_{x,m}\mu+\frac{\partial}{\partial y}H_{y,m}\mu+\frac{\partial}{\partial z}H_{z,m}\mu=0,\nonumber\\
 &-k_{x,m}E_{x,m}\epsilon+\frac{\partial}{\partial y}E_{y,m}\epsilon+\frac{\partial}{\partial z}E_{z,m}\epsilon=\rho_m.
\end{align}
Hence we have reduced the 3D problem, Eqs.~(\ref{Eq01}), to an infinite set of independent 2D problems, Eqs.~(\ref{Eq05}). Our approach is similar to one used in~\cite{Paech07} for solving Maxwell's equations in the frequency domain. 
	
%
\section{Longitudinal wake as a harmonic function}\label{sec:3}
%

Let us consider a line-charge beam with vanishing transverse dimensions,
\begin{align}\label{Eq07}
	\rho(x_0,y_0,x,y,s)&=Q\delta(x-x_0)\delta(y-y_0)\lambda(s),
	\nonumber\\
	j_z(x_0,y_0,x,y,s)&=c\rho(x_0,y_0,x,y,s),\qquad
\end{align}
where $x_0$, $y_0$, define the transverse offset of the beam, ${s=z-ct}$  is the local longitudinal coordinate in the bunch, $Q$ is the bunch charge and $\lambda(s)$ is the longitudinal bunch profile [for a point charge, $\lambda(s) = \delta(s)$]. The longitudinal wake potential $W_{\parallel}$ at point $(x,y,s)$ is defined as~\cite{Zotter98} 
    \begin{equation}\label{Eq08}
    W_{\parallel}(x_0,y_0,x,y,s)
    =
    Q^{-1}\int_{-\infty}^{\infty}[E_z(x,y,z,t)]_{t=({z-s})/{c}}dz,
    \end{equation}
where the electric field on the right-hand side is the solution to Maxwell's equation with the sources of Eqs.~\eqref{Eq07} (this field, of course, is also a function of $x_0$ and $y_0$ omitted in the arguments of $E_z$ for brevity).

It is well known~\cite{Wei92} that the longitudinal wake potential satisfies the Laplace equation for the coordinates $x$, $y$, of the witness particle
    \begin{equation}\label{Eq12} 
    	\left(\frac{\partial^2}{\partial x^2}
	+
	\frac{\partial^2}{\partial y^2}\right) W_{\parallel}(x_0,y_0,x,y,s)
	=0.
    \end{equation}

In order to express the wake potential in structures of constant width through one-dimensional functions we need to prove that the longitudinal wake potential satisfies also the Laplace equation   with respect to the coordinates $x_0$, $y_0$,   of the source particle. This statement will actually be proven below for arbitrary structures without any restrictions on the structure geometry. The only requirement imposed on the system is that the incoming and outgoing waveguides have perfectly conducting walls.

The proof is based on the directional symmetry relation between the wake potential $W_{\parallel}$ for the case when the particle travels through the system in the positive $z$ direction, and the wake potential  $ W_{\parallel}^{(-)}$ corresponding to the case when the particle travels through the same system but in the negative $z$ direction.
It was shown in Ref.~\cite{Zag06} that the two wake potentials are related by 
    \begin{align}\label{Eq15}
    W_{\parallel}^{(-)}(\vec r_2,\vec r_1,s)
    -
    W_{\parallel}(\vec r_1,\vec r_2,s)=2cZ_e(\vec r_1,\vec r_2)\lambda(s),
    \end{align}
where    
    \begin{align}\label{Eq15-1}
    Z_e(\vec r_1,\vec r_2)
    =
    \frac{\epsilon_0}{c}
    \left(\int_{S_A}
    ds\,\nabla\phi_A(\vec r_1,\vec r)\cdot \nabla\phi_A(\vec r_2,\vec r) 
    -
    \int_{S_B}
    ds\, \nabla\phi_B(\vec r_1,\vec r)\cdot\nabla\phi_B(\vec r_2,\vec r) 
    \right)
    ,
    \end{align}
$\epsilon_0$ is the permittivity of vacuum, $\vec r_1 = (x_0,y_0)$ and $\vec r_2=(x,y)$ are two dimensional vectors, and $\phi_A$, $\phi_B$,  are the Green functions for the Laplacian in the incoming and outgoing pipes with cross-sections, $S_A$, $S_B$:
\begin{align} \label{Eq16}
 \Delta\phi_A(\vec r_i,\vec r)&=-\epsilon_0^{-1}\delta(\vec r-\vec r_i),\qquad
 \vec r \in S_A, \nonumber\\
 \phi_A(\vec r_i,\vec r)&=0,\qquad
 \vec r \in \partial S_A, \nonumber\\
 \Delta\phi_B(\vec r_i,\vec r)&=-\epsilon_0^{-1}\delta(\vec r-\vec r_i),\qquad
 \vec r \in S_B, \nonumber\\
 \phi_B(\vec r_i,\vec r)&=0,\qquad
 \vec r \in \partial S_B, \qquad
 i=1,2,
\end{align}
where $\partial S$  is the boundary of $S$. Note that $\vec r_1$ and $\vec r_2$ are offsets of, respectively, the leading and the trailing  particles in $W_{\parallel}$, while $\vec r_2$ is the offset of the leading particle, and $\vec r_1$ is the offset of the trailing particle in $W_{\parallel}^{(-)}$. Note also that, according to the general property of the wake potentials~\eqref{Eq12}, $W_{\parallel}^{(-)}$ satisfies the Laplace equation
    \begin{equation}\label{Eq12-1} 
    	\left(\frac{\partial^2}{\partial x_0^2}
	+
	\frac{\partial^2}{\partial y_0^2}\right) W_{\parallel}^{(-)}(x,y,x_0,y_0,s)
	=0.
    \end{equation}

With the help of Green's first identity ($\vec n$  is the outward pointing unit vector normal to the line element $dl$),
    \begin{equation*}
    \int_S \nabla\varphi\cdot\nabla\psi ds
    =
    \int_{\partial S} \varphi\partial_{\vec n}\psi dl
    -
    \int_S \varphi\Delta\psi ds
    ,
    \end {equation*}
we can rewrite Eq.~(\ref{Eq15}) as
    \begin{equation}\label{Eq17}
    W_{\parallel}^{(-)}(\vec r_2,\vec r_1,s)-W_{\parallel}(\vec r_1,\vec r_2,s)
    =
    2\lambda(s)\left[ \phi_B(\vec r_1,\vec r_2)-\phi_A(\vec r_1,\vec r_2)\right],
    \end{equation}
where we have used the symmetry of Green's functions, $\phi_A(\vec r_1,\vec r_2)=\phi_A(\vec r_2,\vec r_1)$, $\phi_B(\vec r_1,\vec r_2)=\phi_B(\vec r_2,\vec r_1)$. We now apply the Laplacian operator to the coordinates $x_0$, $y_0$ in Eq.~(\ref{Eq17}) and use relation~\eqref{Eq12-1} to obtain
    \begin{equation}\label{Eq18}
    	\left(\frac{\partial^2}{\partial x_0^2}
	+
	\frac{\partial^2}{\partial y_0^2}\right) W_{\parallel}(x_0,y_0,x,y,s)
	=0.
    \end{equation}
This proves that the wake potential, in addition to being a harmonic function with respect to the coordinates of the trailing particle, is also a harmonic function with respect to the coordinates of the leading particle. To our knowledge, this general property of wakefields has not been previously reported in the literature.

%
\section{Wake potential representation in structures of constant width}\label{sec:4}
%
%
It is well known~\cite{Wei92} that for rotationally symmetric structures the wake potential can be represented through one dimensional functions, with only one function for each azimuthal mode number $m$, 
    \begin{align}\label{Eq09}
    &W_{\parallel}(x_0,y_0,x,y,s)
    =
    \sum\limits_{m=0}^\infty W_m(s)r_0^m r^m\cos(m(\theta-\theta_0)),\nonumber\\
    &x_0=r_0\cos(\theta_0),\qquad y_0=r_0\sin(\theta_0),\qquad
    x=r\cos(\theta),\qquad y=r\sin(\theta).
    \end{align}
As it turns out, a similar, comparably simple representation exists for the wake potential in structures of rectangular cross-section and constant width; however, in the latter case four one-dimensional functions are needed for each mode number $m$.

Substituting  Eq.~(\ref{Eq07}) into~\eqref{Eq03} we find for the  charge distribution 
    \begin{align}\label{Eq10}
    	\rho(x_0,y_0,x,y,s)
	&=\frac{1}{w}\sum\limits_{m=1}^\infty \rho_m(y_0,y,s)\sin(k_{x,m} x_0)\sin(k_{x,m} x),\nonumber\\
    	\rho_m(y_0,y,s)&=Q\delta(y-y_0)\lambda(s).
    \end{align}
It then follows from Maxwell's equations~\eqref{Eq05} that all components of the fields will be proportional to $\sin(k_{x,m} x_0)$ and using Eqs.~(\ref{Eq04}) and (\ref{Eq08}) we find
    \begin{equation}\label{Eq11}
    	W_\parallel(x_0,y_0,x,y,s)
	=
	\frac{1}{w}\sum\limits_{m=1}^\infty W_m(y_0,y,s)\sin(k_{x,m} x_0)\sin(k_{x,m} x).
    \end{equation}
If we insert representation~(\ref{Eq11}) into Eq.~(\ref{Eq12}), we obtain the one-dimesional Helmholz equation
    \begin{equation}\label{Eq13}
    	\frac{\partial^2}{\partial y^2}W_m(y_0,y,s)-(k_{x,m})^2 W_m(y_0,y,s)=0,
    \end{equation}
which has a general solution of the form
    \begin{equation}\label{Eq14}
    	W_m(y_0,y,s)=W_m^c(y_0,s)\cosh(k_{x,m} y)+W_m^s(y_0,s)\sinh(k_{x,m} y).
    \end{equation}
Finally from Eq.~(\ref{Eq18}) it follows that functions $W_m^c(y_0,s)$   and  $W_m^s(y_0,s)$  satisfy equations	
    \begin{align}\label{Eq19}
    	\frac{\partial^2}{\partial y_0^2}W_m^c(y_0,s)-(k_{x,m})^2 W_m^c(y_0,s)&=0,\nonumber\\
    	\frac{\partial^2}{\partial y_0^2}W_m^s(y_0,s)-(k_{x,m})^2 W_m^s(y_0,s)&=0.
    \end{align}
These equations can again be easily integrated with the result:
    \begin{align}\label{Eq21-1}
    	W_{\parallel}(x_0,y_0,x,y,s)
	&=
	\frac{1}{w}\sum\limits_{m=1}^\infty W_m(y_0,y,s)\sin(k_{x,m} x_0)\sin(k_{x,m} x),
    \end{align}
where
    \begin{align}\label{Eq21}
    	W_m(y_0,y,s)&=\left[W_m^{cc}(s)\cosh(k_{x,m} y_0)
	+W_m^{sc}(s)\sinh(k_{x,m} y_0)\right]\cosh(k_{x,m} y)
	\nonumber\\
    	&+\left[W_m^{cs}(s)\cosh(k_{x,m} y_0)+W_m^{ss}(s)\sinh(k_{x,m} 
y_0)\right]\sinh(k_{x,m} y).
    \end{align}
Thus, we have proven that, in structures of constant width, for each mode number $m$ four functions are needed to completely describe the longitudinal wake potential. These functions can be calculated as follows
    \begin{align}\label{Eq22}
    W_m^{cc}&=W_m(0,0,s),\qquad
    &W_m^{sc}&=\frac{1}{k_{x,m}}\frac{\partial}{\partial x}W_m(0,0,s),\nonumber\\
    W_m^{cs}&=\frac{1}{k_{x,m}}\frac{\partial}{\partial x_0}W_m(0,0,s),\qquad
    &W_m^{ss}&=\frac{1}{(k_{x,m})^2}\frac{\partial^2}{\partial x\partial x_0}W_m(0,0,s),
    \end{align}
where the $m^{\rm th}$ modal component of the wake potential
    \begin{equation}\label{Eq23}
    W_m(y_0,y,s)
    =
    Q^{-1}\int_{-\infty}^{\infty}[E_{z,m}(y,z,t)]_{t=({z-s})/{c}}dz
    \end{equation}
is excited by a charge distribution that does not depend on $x$, 
    \begin{equation}\label{Eq24}
    	\rho_m(y_0,y,s)=Q\delta(y-y_0)\lambda(s).
    \end{equation}

With a knowledge of the longitudinal wake we can calculate the transverse wakes. For example, the vertical wake potential, $W_y$, can be easily found through the Panofsky-Wenzel theorem~\cite{Wei92}
    \begin{equation}\label{Eq25}
    	\frac{\partial}{\partial s} W_{y}(x_0,y_0,x,y,s)=
    	\frac{\partial}{\partial y} W_{\parallel}(x_0,y_0,x,y,s).
    \end{equation}
In the next section we will analyze both the longitudinal and the transverse wakes assuming that the structure under consideration has a symmetry axis.

%
\section{Structures with horizontal symmetry plane}\label{sec:5}
%

Let us consider a structure of constant width $2w$ that also has a vertical symmetry plane, at  $y=0$. Structures in Fig.~\ref{Fig01} (a) and (b) possess this symmetry; hence, they have a symmetry axis located at $x=w$, $y=0$. Due to the symmetry, the wake potential satisfies the equation
    \begin{equation}\label{Eq26}
    	W_{\parallel}(x_0,y_0,x,y,s)
	=
	W_{\parallel}(x_0,-y_0,x,-y,s),
    \end{equation}
and Eq.~(\ref{Eq21}) simplifies:
    \begin{align}\label{Eq27}
    	W_m(y_0,y,s)&=W_m^{cc}(s)\cosh(k_{x,m} y_0)\cosh(k_{x,m} y)
    	+W_m^{ss}(s)\sinh(k_{x,m} y_0)\sinh(k_{x,m} y).
    \end{align}
Note that
    \begin{align}\label{Eq27-1}
     	W_m(y_0,y,s)=W_m(y,y_0,s).
    \end{align}

Let us consider the transverse wakes in such structures. We first introduce the integrated wake functions (sometimes called the step function response)
    \begin{equation}\label{Eq29}
     S_m^{cc}=\int_{-\infty}^s W_m^{cc}(s')ds',\quad
     S_m^{ss}=\int_{-\infty}^s W_m^{ss}(s')ds'.\quad
    \end{equation}
It then follows from \eqref{Eq25} that the transverse wake function can be written as
    \begin{equation}\label{Eq30}
    	W_y(x_0,y_0,x,y,s)
	=
	\frac{1}{w}\sum\limits_{m=1}^\infty k_{x,m} W_{y,m}(y_0,y,s)\sin(k_{x,m} x_0)\sin(k_{x,m} x),
    \end{equation}
where
    \begin{align}\label{Eq31}
    	W_{y,m}(y_0,y,s)
	&=S_m^{cc}(s)\cosh(k_{x,m} y_0)\sinh(k_{x,m} 
y)+S_m^{ss}(s)\sinh(k_{x,m} y_0)\cosh(k_{x,m} y),
    \nonumber\\
    	W_x(x_0,y_0,x,y,s)
	&=
	\frac{1}{w}\sum\limits_{m=1}^\infty k_{x,m} W_{x,m}(y_0,y,s)\sin(k_{x,m} x_0)\cos(k_{x,m} x),
    \\	
    W_{x,m}(y_0,y,s)
    &=S_m^{cc}(s)\cosh(k_{x,m} y_0)\cosh(k_{x,m} y)+S_m^{ss}(s)\sinh(k_{x,m} 
y_0)\sinh(k_{x,m} y).
    \nonumber
	\end{align}
Representations (\ref{Eq30}), (\ref{Eq31}), are valid for arbitrary offsets of leading  and trailing particles. 

For small offsets near the symmetry axis, $x=w$, $y=0$, the transverse wake potential is usually expanded in Taylor series, 
    \begin{align}\label{Eq32}
    	W_y(w,y_0,w,y,s)
	&\approx
	y_0\frac{\partial}{\partial y_0}W_y(w,y_0,w,0,s)\big|_{y_0=0}
	+y\frac{\partial}{\partial y}W_y(w,0,w,y,s)\big|_{y=0}.
    \end{align}
The first term in~(\ref{Eq32}) is usually called the transverse {\it dipole} wake in  the $y$-direction. It can be calculated as follows
	\begin{equation}\label{Eq33}
	W_{y,d}(s)\equiv\frac{\partial}{\partial y_0}W_y(w,y_0,w,0,s)\big|_{y_0=0}
	= 
	\frac{1}{w}\sum\limits_{m=1,\mathrm{odd}}^{\infty}(k_{x,m})^2 S_m^{ss}(s).
	\end{equation}
The second term in~(\ref{Eq32}) is called the transverse {\it quadrupole} wake in  $y$-direction; it is obtained by
	\begin{equation}\label{Eq34}
	W_{y,q}(s)\equiv\frac{\partial}{\partial y}W_x(w,0,w,y,s)\big|_{y=0}
	= \frac{1}{w}\sum\limits_{m=1,\mathrm{odd}}^{\infty}(k_{x,m})^2 
S_m^{cc}(s).
	\end{equation}
The transverse wakes in the $x$ direction are obtained by equations corresponding to those of Eqs.~(\ref{Eq33}), (\ref{Eq34}). Note that $W_{y,q}(s)=-W_{x,q}(s)$.

In numerical calculations of structures with symmetry we can use the approach of paper~\cite{Zag02} that allows us to reduce the calculation domain in half. Indeed the charge distribution (\ref{Eq24}) can be written as a sum of symmetric and antisymmetric parts
	\begin{align}
	\rho_m(y_0,y,s)&=\rho_m^E(y_0,y,s)+\rho_m^H(y_0,y,s),\label{Eq35}
	\end{align}
where
	\begin{align}
	\rho_m^H(y_0,y,s)&=\frac{1}{2}Q \left[\delta(y-y_0)+\delta(y+y_0) \right] \lambda(s),\label{Eq36}\\
	\rho_m^E(y_0,y,s)&=\frac{1}{2}Q\left[\delta(y-y_0)-\delta(y+y_0) \right]\lambda(s).\label{Eq37}
	\end{align}
In problems with the symmetric driving charges (\ref{Eq36}), the tangential component of the magnetic field will be zero in the symmetry plane  (the so called ``magnetic'' boundary condition). In problems with the antisymmetric driving charges (\ref{Eq37}) the tangential component of the electric field will be zero in the symmetry plane  (the ``electric'' boundary condition). Thus, instead of solving the system of equations (\ref{Eq05}) in the whole domain, one can solve two independent problems in half of the domain: one problem with the ``magnetic'' boundary condition at $y=0$ and one problem with the ``electric'' boundary condition at $y=0$. This is true not only for the line-charge current distribution~(\ref{Eq07}), but for any arbitrary three dimensional charge distribution $\rho(x,y,z,t)$. From solutions $W_m^H(y_0,y,s)$  and $W_m^E(y_0,y,s)$   of the two problems we can easily find the one dimensional modal functions in Eq. (\ref{Eq27}):
	\begin{equation}\label{Eq38}
	W_m^{cc}(s)=W_m^H(0,0,s),\qquad 
	W_m^{ss}(s)=(k_{x,m})^{-2}\frac{\partial^2}{\partial y_0\partial y}W_m^E(y_0,y,s)
	\big|_{y,y_0=0}.
	\end{equation}

%
\section{Symmetry relations for the transverse wake potential}\label{sec:6}
%

In the general case, from the directional symmetry relation, Eqs.~\eqref{Eq15}, discussed in Section~\ref{sec:3}, one cannot immediately infer the corresponding symmetry for the transverse wakes taken in the positive and negative directions. It turns out, however, that  for structures of constant width with a symmetry axis, such  a directional symmetry relation for the transverse wake potential can be proven. Indeed, from the symmetry relation (\ref{Eq27-1}) and Eq.~(\ref{Eq21-1}) follows the symmetry of the longitudinal wake potential,
	\begin{equation}\label{Eq39}
	W_\parallel(x_0,y_0,x,y,s)=W_\parallel(x,y,x_0,y_0,s).
	\end{equation}
Hence the directional symmetry relation~(\ref{Eq17}) for the longitudinal wake potential can now be written as 		 
    \begin{equation}\label{Eq40}
    W_{\parallel}^{(-)}(\vec r_1,\vec r_2,s)-W_{\parallel}(\vec r_1,\vec r_2,s)
    =2\lambda(s)\left[ \phi_B(\vec r_1,\vec r_2)-\phi_A(\vec r_1,\vec r_2)\right];
    \end{equation}
and from the Panofsky-Wenzel theorem~(\ref{Eq25}) it follows that  
    \begin{align}\label{Eq41}
    \vec W_{\perp}^{(-)}(\vec r_1,\vec r_2,s)
    -
    \vec W_{\perp}(\vec r_1,\vec r_2,s)
    &=2\Lambda(s)
    \left[\nabla\phi_B(\vec r_1,\vec r_2)
    -\nabla\phi_A(\vec r_1,\vec r_2)\right],\nonumber\\
    \Lambda(s)
    &=\int_{-\infty}^s\lambda(s')ds'.
    \end{align}
In the last equation we have used the vectorial notation for the transverse wake, $\vec W_{\perp} = ( W_{x},W_{y})$.

Let us now calculate the directional relation for the dipole and quadrupole wakes defined by Eqs.~(\ref{Eq33}), (\ref{Eq34}). Denote the vertical aperture of  the incoming rectangular pipe by $2g_A$  and that of the outgoing  pipe $2g_B$. (Both pipes have the same width $2w$.) The Green function for Poisson's equation in a rectangular pipe, $\phi_A(x_0,y_0,x,y)$, has been obtained by Gluckstern, {\it et al}~\cite{Gluckstern}. For $x_0=w$, their result reads:\footnote{There is a typo in Eq.~(5.11) of Ref.~\cite{Gluckstern} that is corrected in~\eqref{rectangular_trans_eq}.}
\begin{align}\label{rectangular_trans_eq}
\phi_A(w,y_0,x,y)&=-\frac{1}{\pi\epsilon_0}\sum_{n=1}^\infty\frac{e^{-\frac{n\pi w}{2g_A}}\cosh
    \frac{n\pi (x-w)}{2g}}{n\cosh \frac{n\pi
    w}{2g_A}}\sin\frac{n\pi}{2g}(y+g_A)\sin\frac{n\pi}{2g}(y_0+g_A)\nonumber\\
    &+\ln\left\{\left[(x-w)^2+(y-y_0)^2\right] \frac{\sinh^2\frac{\pi
    (x-w)}{4g_A}+\cos^2\frac{\pi}{4g_A}(y+y_0)}{\sinh^2\frac{\pi
    (x-w)}{4g_A}+\sin^2\frac{\pi }{4g_A}(y-y_0)}\right\}\nonumber\\
    &-\ln\left[{(x-w)^2+(y-y_0)^2}\right].
    \end{align}
The last term is singular in the limit $x\to w$ and $y\to y_0$. All other terms are
finite for all admissible $x$ and $y$, and the sum in the first term rapidly converges. The Green's function of the outgoing pipe, $\phi_B(w,y_0,x,y)$, is obtained from~\eqref{rectangular_trans_eq} by replacing $g_A \to g_B$. It is then easy to see that the singular terms cancel in the difference $\phi_A-\phi_B$; hence this function is finite within the domain of its definition. If we follow the approach of paper~\cite{Bane07} and use Eqs.~(39), (40), (45) from that paper, then from Eq.~(\ref{Eq41}) our final result reads
    \begin{align}\label{direct_transv}
    W_{y,d}^{(-)}(s)-W_{y,d}(s)
    &=
    2c\Lambda(s)\left[Z_{y,d}(g_A)-Z_{y,d}(g_B)\right],\nonumber\\
    W_{y,q}^{(-)}(s)-W_{y,q}(s)
    &=
    2c\Lambda(s)\left [Z_{y,q}(g_A)-Z_{y,q}(g_B)\right ],\nonumber\\
    Z_{y,d}(g)
    &=
    \frac{\pi Z_0}{12g^2}\left[1+24\sum_{m=1}^\infty\frac{m}{1+e^{2\pi m (w/g)}}\right],\nonumber\\
    Z_{y,q}(g)
    &=
    \frac{\pi Z_0}{24g^2}\left[1-24\sum_{m=1}^\infty\frac{2m-1}{1+e^{\pi(2m-1)(w/g)}}\right].
    \end{align}

%
\section{TE/TM scheme in matrix notation}\label{sec:7}
%

In this section we describe a particular realization of the implicit TE/TM scheme introduced in~\cite{Zag05}. The scheme will be discussed in the context of the Finite Integration Technique~\cite{Wei96}.
\begin{figure}[htbp]
\centering
\includegraphics*[height=50mm]{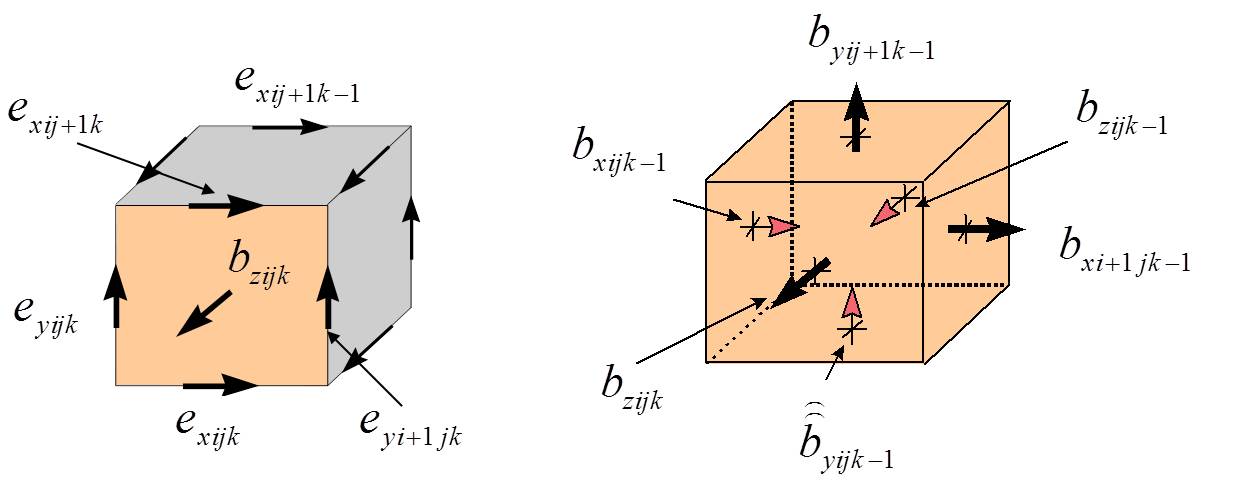}
\caption{One cell of cell complex $G$ showing the positions of the voltage and 
magnetic flux components}\label{Fig03}
\end{figure}

We will consider Maxwell's equations in their integral form on a domain $\Omega$ with the linear non-dispersive constitutive relations:
\begin{align}\label{ME_in IF}
\oint\limits_{\partial S}\vec{E} d\vec{l}=-\frac{\partial}{\partial 
t}\oiint\limits_{S}\vec{B} d\vec{s},\quad
\oint\limits_{\partial S}\vec{H} d\vec{l}=\frac{\partial}{\partial t}\oiint\limits_{S}\vec{D} d\vec{s}+\oiint\limits_{S}\vec{j} d\vec{s},\qquad
\forall S\subset\Omega,\nonumber\\
\oiint\limits_{\partial V}\vec{D} d\vec{s}=\oiiint\limits_{V}\rho dv,\qquad
\oiint\limits_{\partial V}\vec{B} d\vec{s}=0,\qquad
\forall V\subset\Omega,\nonumber\\
\vec{D}=\epsilon\vec{E},\qquad \vec{B}=\mu\vec{H},\qquad
\forall x\in\Omega.
\end{align}	

Let us start by introducing a grid-based decomposition of the entire computation domain $\Omega$ into cell complex $G$. We use here a three dimensional Cartesian mesh in $x,y,z$  coordinates, with the corresponding indexing  $i,j,k$. Unlike in finite-difference methods where one starts by allocating the field components, we begin by taking the voltage along cell edges and the magnetic flux through cell facets as the computational unknowns:
\begin{align}
e_{\vartheta}=\int_{L_{\vartheta}}{\vec{E}d\vec{l}},\quad
b_{\vartheta}=\iint_{{S}_{\vartheta}}{\vec{B}d\vec{s}},
\end{align}
where $\vartheta$  is a mesh multi-index and  $L_{\vartheta},S_{\vartheta}\in G$.  Solving Faraday's law in integral form for the front surface shown in Fig.~{\ref{Fig03}} yields:
\begin{equation*} -e_{x,i,j,k}+e_{y,i,j,k}+e_{x,i,j+1,k}-e_{y,i+1,j,k}=-\frac{\partial}{\partial t}b_{z,i,j,k}. 
\end{equation*}
Note, that this representation is still exact, as $e_{\vartheta}$ is (by definition) the exact electric voltage along one edge of the cell, and similarly $b_{\vartheta}$   represents the exact value of the magnetic flux density integral over the cell surface. If we compose column vectors
\begin{align} 
\vec{e}=
\begin{pmatrix}
\vec{e}_x\\\vec{e}_y\\\vec{e}_z
\end{pmatrix},\qquad
\vec{b}=
\begin{pmatrix}
\vec{b}_x\\\vec{b}_y\\\vec{b}_z
\end{pmatrix}
\end{align} 
from all voltage and flux components, we can write the combination of all equations over all surfaces in an elegant matrix form as
\begin{equation}\label{}
\mathbf{Ce}=-\frac{\partial}{\partial t}\mathbf b.
\end{equation}
The matrix $\mathbf{C}$ picks the affected components out of the long vector to make up the corresponding equation. $\mathbf{C}$ is thus the discrete curl operator over the mesh $G$. With an appropriate indexing scheme the curl matrix has a 3x3 block structure:
\begin{align}
\mathbf{C}&=
\begin{pmatrix}
\mathbf{0}&-\mathbf{P}_z&\mathbf{P}_y\\
\mathbf{P}_z&\mathbf{0}&-\mathbf{P}_x\\
-\mathbf{P}_y&\mathbf{P}_x&\mathbf{0}
\end{pmatrix}.
\end{align}
The double-banded, topological $\mathbf{P}_x,\mathbf{P}_y,\mathbf{P}_z$-matrices take the role of discrete partial differential operators.

The second important differential operator in Maxwell's equations (\ref{Eq01}) is the divergence operator. In order to construct a discrete divergence operator we integrate Maxwell's equation $\oiint\limits_{\partial V}\vec{B} d\vec{s}=0$  over the entire surface of a mesh cell depicted in Fig.~\ref{Fig03}. By adding up the six fluxes for each cell and by writing down all such equations for the entire mesh we obtain a discrete analogue to the divergence equation:
	\begin{align}\label{}
\mathbf{Sb}=\mathbf{0},\qquad
\mathbf{S}=\left(\mathbf{P}_x\quad\mathbf{P}_y\quad\mathbf{P}_z\right).
\end{align}

The discretization of the remaining Maxwell equations requires the introduction of a second cell complex $\tilde{G}$ which is dual to the primary cell complex $G$. For the Cartesian grid the dual complex $\tilde{G}$ is defined by taking the foci of the cells of $G$ as gridpoints for the mesh cells of $\tilde{G}$. 
We again introduce the computational unknowns as integrals
\begin{align}
&h_{\vartheta}=\int_{\tilde{L}_{\vartheta}}{\vec{H}d\vec{l}},\quad
d_{\vartheta}=\iint_{\tilde{S}_{\vartheta}}{\vec{D}d\vec{s}},\quad
j_{\vartheta}=\iint_{\tilde{S}_{\vartheta}}{\vec{j}d\vec{s}},
\end{align}
where $\vartheta$  is a mesh multi-index and $\tilde{L}_{\vartheta},\tilde{S}_{\vartheta}\in \tilde{G}$.

Following an equivalent procedure for the remaining Maxwell's equations, but using the dual cell complex $\tilde{G}$,  we obtain a set of four discrete equations representing Maxwell's equations on a dual grid: 
\begin{align}\label{Eq42}
\mathbf{Ce}&=-\frac{\partial}{\partial t}\mathbf b,\qquad
&\mathbf{C^{*}h}&=\frac{\partial}{\partial t}\mathbf d+\mathbf j,\nonumber\\
\mathbf{Sb}&=\mathbf{0},\qquad
&\mathbf{\tilde{S}d}&=\mathbf{q},\qquad
\mathbf{e,b}\in G,\qquad
\mathbf{h,d}\in \tilde G,\nonumber\\
\mathbf{\tilde{S}}&=\left(-\mathbf{P}_x^{*}\quad-\mathbf{P}_y^{*}\quad-\mathbf{P}_z^{*}\right),
\end{align}
where the asterisk denotes the Hermetian adjoint operator.

Equations (\ref{Eq42}) are completed by the discrete form of the material relations (constitutive equations) which appear (in the simplest linear case) as matrix equations
\begin{align}\label{Eq_material}
\mathbf{e}&=c\mathbf{M}_{\epsilon^{-1}}\mathbf d,\qquad
&\mathbf{h}&=c\mathbf{M}_{\mu^{-1}}\mathbf b.
\end{align}
with $\mathbf{M}_{\epsilon^{-1}}$ the discrete inverse permittivity matrix,  and $\mathbf{M}_{\mu^{-1}}$ the inverse permeability matrix.

\begin{figure}[htbp]
\centering
\includegraphics*[height=50mm]{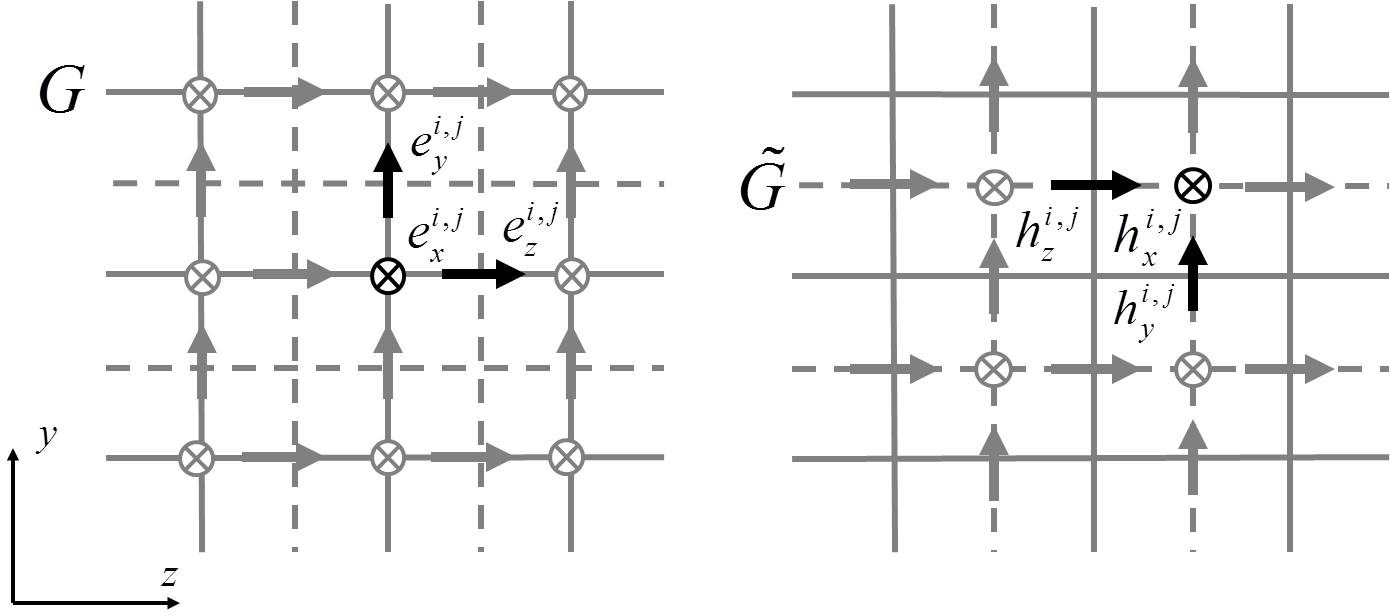}
\caption{Grid doublet $(G,\tilde G)$ in $(y,z)$-plane.}\label{Fig04}
\end{figure}

\begin{figure}[htbp]
\centering
\includegraphics*[height=50mm]{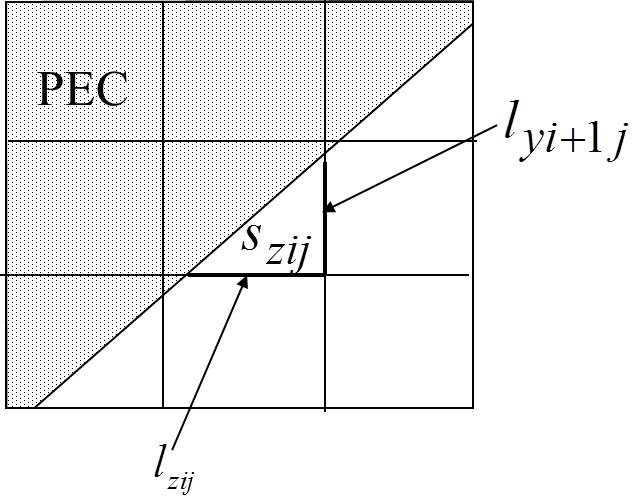}
\caption{Curved PEC-boundary in a Cartesian mesh.}\label{Fig04b}
\end{figure}

In the case of Cartesian grids (or, more generally, whenever the primary and dual grids are orthogonal) all material operators can be represented by diagonal matrices. Note that the material matrices contain both averaged material parameters, and the lengths and areas of the grid edges and faces, respectively. In the standard staircase approximation the material matrices contain elements (without double indices for simplicity of notation)
\begin{align}
\epsilon_{pijk}^{^-1}=(c\epsilon)^{-1}\frac{L_{pijk}}{\tilde{S}_{pijk}},\qquad
\mu_{pijk}^{^-1}=(c\mu)^{-1}\frac{S_{pijk}}{\tilde{L}_{pijk}},\qquad
\end{align}
with $p=x,y,z,$  and the face areas and edge lengths of the primary and secondary grid given by $S,L,\tilde{S},\tilde{L}$, respectively.

In the conformal scheme~\cite{Thoma97}, we allow the cells of the computational grid to be only partially filled by a perfectly electric conducting (PEC) material and with an arbitrarily shaped interface. To model such a case we modify only the elements of the material matrices:
\begin{align}
\epsilon_{pijk}^{^-1}=(c\epsilon)^{-1}\frac{l_{pijk}}{\tilde{S}_{pijk}},\qquad
\mu_{pijk}^{^-1}=(c\mu)^{-1}\frac{s_{pijk}}{\tilde{L}_{pijk}},\qquad
\end{align}
where $s,l$   denote the reduced cell areas and lengths, including only those parts inside the computational domain (outside PEC material), as shown in Fig.~\ref{Fig04b}.
	
The complete set of equations (\ref{Eq42}) and (\ref{Eq_material}) is referred to as Maxwell's grid equations. 
	
For our 2D rectangular structures, Maxwell's equations~(\ref{Eq01}) reduce to the modal equations (\ref{Eq05}). In addition, the dual three-dimensional cell complex $(G,\tilde{G})$ reduces to the two-dimensional plane complex shown in Fig.~\ref{Fig04}. In Maxwell's grid equations the discrete operator $\mathbf{P}_x$ reduces to the diagonal matrix $\mathbf{k}_{x,m}=k_{x,m} \mathbf{I}$, where $\mathbf{I}$  is the unit matrix. 
 
System (\ref{Eq42}) is a time-continuous and space-discrete approximation of problem (\ref{Eq05}). The next step is a discretization of the system in time. The field components can be split in time and the leap-frog method can be applied. With ``electric/magnetic" splitting, a well known Yee's scheme~\cite{Yee66} will be obtained. However Yee's scheme has large dispersion errors along the grid lines. In the following we consider an alternative TE/TM splitting scheme~\cite{Zag05}, one that does not have dispersion errors in the longitudinal direction. 

Let us rewrite Eqs.~(\ref{Eq42}) in an equivalent form
\begin{equation}\label{Eq43}
\frac{\partial}{\partial \tau}\mathbf{u}=\mathbf{M_u(T_u u+Lv-j_u)},\qquad
\frac{\partial}{\partial \tau}\mathbf{v}=\mathbf{M_v(T_v v-L^{*}u-j_v)},
\end{equation}
where $\tau=ct$ and
\begin{align}
\mathbf{T_u}&=
\begin{pmatrix}
\mathbf{0}&\mathbf{0}&-\mathbf{P}_y\\
\mathbf{0}&\mathbf{0}&\mathbf{k}_{x,m}\\
\mathbf{P}_y^{*}&-\mathbf{k}_{x,m}&\mathbf{0}
\end{pmatrix},\quad
\mathbf{T_v}=
\begin{pmatrix}
\mathbf{0}&\mathbf{0}&-\mathbf{P}_y^{*}\\
\mathbf{0}&\mathbf{0}&\mathbf{k}_{x,m}\\
\mathbf{P}_y&-\mathbf{k}_{x,m}&\mathbf{0}
\end{pmatrix},\quad
\mathbf{L}=
\begin{pmatrix}
\mathbf{0}&\mathbf{P}_z&\mathbf{0}\\
-\mathbf{P}_z&\mathbf{0}&\mathbf{0}\\
\mathbf{0}&\mathbf{0}&\mathbf{0}
\end{pmatrix},\nonumber\\
\mathbf{M_u}&=
\begin{pmatrix}
\mathbf{M}_{\mu_x^{-1}}&\mathbf{0}&\mathbf{0}\\
\mathbf{0}&\mathbf{M}_{\mu_y^{-1}}&\mathbf{0}\\
\mathbf{0}&\mathbf{0}&\mathbf{M}_{\epsilon_z^{-1}}
\end{pmatrix},\quad
\mathbf{M_v}=
\begin{pmatrix}
\mathbf{M}_{\epsilon_x^{-1}}&\mathbf{0}&\mathbf{0}\\
\mathbf{0}&\mathbf{M}_{\epsilon_y^{-1}}&\mathbf{0}\\
\mathbf{0}&\mathbf{0}&\mathbf{M}_{\mu_z^{-1}}
\end{pmatrix},\nonumber\\
\mathbf{u}&=
\begin{pmatrix}
\mathbf{h}_x\\\mathbf{h}_y\\\mathbf{e}_z
\end{pmatrix},\quad
\mathbf{v}=
\begin{pmatrix}
\mathbf{e}_x\\\mathbf{e}_y\\\mathbf{h}_z
\end{pmatrix},\quad
\mathbf{j_u}=
\begin{pmatrix}
\mathbf{0}\\\mathbf{0}\\\mathbf{j}_z
\end{pmatrix},\quad
\mathbf{j_v}=
\begin{pmatrix}
\mathbf{j}_x\\\mathbf{j}_y\\\mathbf{0}
\end{pmatrix}.
\end{align}

Applying TE/TM splitting~\cite{Zag05} of the field components in time to system (\ref{Eq43}), the following numerical scheme is obtained
\begin{align}\label{Eq44}
\frac{\mathbf{u}^{n+0.5}-\mathbf{u}^{n-0.5}}{\Delta\tau}&= \mathbf{M_u} \left( \mathbf{T_u}\frac{\mathbf{u}^{n+0.5}+\mathbf{u}^{n-0.5}}{2} +\mathbf{Lv}^n-\mathbf{j_u}^n\right),\nonumber\\
\frac{\mathbf{v}^{n+1}-\mathbf{v}^{n}}{\Delta\tau}&= \mathbf{M_v} \left( \mathbf{T_v}\frac{\mathbf{v}^{n+1}+\mathbf{v}^{n}}{2} -\mathbf{L}^{*}\mathbf{u}^{n+0.5}-\mathbf{j_v}^{n+0.5}\right).
\end{align}
Two-layer operator-difference scheme (\ref{Eq44}) acquires the canonical form~\cite{Sam2001}
\begin{equation}\label{Eq45}
\mathbf{B}\frac{\mathbf{y}^{n+1}-\mathbf{y}^n}{\Delta\tau}+\mathbf{A}\mathbf{y}^{n}=\mathbf{f}^n,
\end{equation}	
where
\begin{align*}
\mathbf{B}&=
\begin{pmatrix}
\mathbf{M_u}^{-1}-\alpha\mathbf{T_u}&\mathbf{0}\\
2\alpha\mathbf{L}^{*}&\mathbf{M_v}^{-1}-\alpha\mathbf{T_v}&\\
\end{pmatrix},\quad
\mathbf{A}=
\begin{pmatrix}
-\mathbf{T_u}&-\mathbf{L}\\
\mathbf{L}^{*}&-\mathbf{T_v}&\\
\end{pmatrix},\\
\mathbf{y}^n&=
\begin{pmatrix}
\mathbf{u}^{n-0.5}\\\mathbf{v}^n
\end{pmatrix},\quad
\mathbf{f}^n=
\begin{pmatrix}
\mathbf{j}_u^n\\\mathbf{j}_v^{n+0.5}
\end{pmatrix},\quad
\alpha=0.5\Delta\tau.
\end{align*} 
It was shown in~\cite{Zag05} that stability of the method is insured if  
\begin{equation}\label{Eq46}
\mathbf{Q\equiv B-\alpha A >0}.
\end{equation}
If the matrix $\mathbf{Q}$   is positive definite, then we can define a discrete energy as 
\begin{equation}
W^n=0.5\langle\mathbf{Qy}^n,\mathbf{y}^n\rangle
\end{equation}
and the discrete energy conservation law holds
\begin{align}\label{Eq47}
\frac{W^{n+1}-W^n}{\Delta\tau}&=-\langle\mathbf{\hat{e}}^{n+0.5},\mathbf{\hat{j}}^{n+0.5}\rangle,\nonumber\\
\mathbf{\hat{e}}^{n+0.5}&=0.5
\begin{pmatrix}
\mathbf{e}_x^{n+1}+\mathbf{e}_x^{n}\\\mathbf{e}_y^{n+1}+\mathbf{e}_y^{n}\\\mathbf{e}_z^{n+0.5}+\mathbf{e}_x^{n-0.5}
\end{pmatrix},
\mathbf{\hat{j}}^{n+0.5}=
\begin{pmatrix}
\mathbf{j}_x^{n+0.5}\\\mathbf{j}_y^{n+0.5}\\\mathbf{j}_z^{n}
\end{pmatrix}.
\end{align}
The stability condition (\ref{Eq46}) in the staircase approximation (of the boundary in vacuum) can be rewritten in the form
\begin{equation}\label{Eq48}
\mathbf{I}-\alpha^2 (\Delta z)^{-2}\,\mathbf{P}_z \mathbf{P}_z^{*}>0.
\end{equation}
This last condition resembles the well-known stability condition of the explicit FDTD scheme for one-dimensional problems. The maximal eigenvalue  of the discrete operator $\mathbf{P}_z \mathbf{P}_z^{*}$  fulfills the relation $\lambda_{max}<4$, and the stability condition of the scheme in the staircase approximation of the boundary in vacuum reads
\begin{equation}\label{Eq49}
\Delta\tau\leq\Delta z. 
\end{equation}
On an equidistant mesh the implicit scheme (\ref{Eq44}) has a second order local approximation error in homogeneous regions.

Relation (\ref{Eq49}) does not contain information about the transverse mesh. Hence the transverse mesh can be chosen independent of stability considerations. 
Following the conventional procedure~\cite{Zag05}, the dispersion relation can be obtained in the form
\begin{equation*}
\left(\frac{\sin\Omega}{\Delta\tau}\right)^2=\left(\frac{\sin K_z}{\Delta z}\right)^2+\left(\frac{\sin K_y}{\Delta y}\cos\Omega\right)^2,
\end{equation*}
where $\Omega=0.5\omega\Delta\tau$, $K_y=0.5k_y\Delta y$, $K_z=0.5k_z\Delta x$. With the ``magic" time step $\Delta\tau=\Delta z$, the scheme does not have dispersion in the longitudinal direction.

Following the approach of~\cite{Zag05} it is easy to show that charge conservation holds:
\begin{align*}
\frac{\mathbf{g}_e^{n+0.5}-\mathbf{g}_e^{n-0.5}}{\Delta\tau}+\mathbf{\tilde{S}\bar{j}}^n=\mathbf{0},\qquad
\frac{\mathbf{g}_h^{n+1}-\mathbf{g}_h^{n}}{\Delta\tau}=\mathbf{0},\nonumber\\
\mathbf{g}_e^{n+0.5}=\mathbf{\tilde{S}}\mathbf{M}_{\epsilon^{-1}}^{-1}\mathbf{\bar{e}}_z^{n+0.5},\quad
\mathbf{g}_h^{n}=\mathbf{S}\mathbf{M}_{\mu^{-1}}^{-1}\mathbf{\bar{h}}_z^{n},
\end{align*}
where functions with overbars are defined as
\begin{equation*}
\mathbf{\bar{f}}^{n}=
\begin{pmatrix}
0.5(\mathbf{f}_x^{n+1}+\mathbf{f}_x^{n}),&
0.5(\mathbf{f}_y^{n+1}+\mathbf{f}_y^{n}),&
\mathbf{f}_z^{n+0.5}
\end{pmatrix}^T.
\end{equation*}

%
\section{TE/TM scheme in component notation}\label{sec:8}
%

In this section we rewrite scheme (\ref{Eq44}) in component notation and discuss the algorithm of numerical solution. 

The field components $\mathbf{h}_x^{n+0.5},\mathbf{h}_y^{n+0.5},\mathbf{e}_z^{n+0.5}$ at time level $n+0.5$ can be found as
\begin{align}\label{Eq50}
\mathbf{\tilde{h}}_x^n&=\mathbf{h}_x^{n-0.5}+\alpha\mathbf{M}_{\mu_x^{-1}}\left[\mathbf{P}_z\mathbf{e}_y^n-\mathbf{P}_y\mathbf{e}_z^{n-0.5}\right],\nonumber\\
\mathbf{\tilde{h}}_y^n&=\mathbf{h}_y^{n-0.5}+\alpha\mathbf{M}_{\mu_y^{-1}}\left[-\mathbf{P}_z\mathbf{e}_x^n+k_{x,m}\mathbf{e}_z^{n-0.5}\right],\nonumber\\
\mathbf{e}_z^{n+0.5}&=\mathbf{e}_z^{n-0.5}+2\alpha\mathbf{W}_e^{-1}\mathbf{M}_{\epsilon_z^{-1}}\left[-k_{x,m}\mathbf{\tilde{h}}_y^n+\mathbf{P}_y^{*}\mathbf{\tilde{h}}_x^n-\mathbf{j}_z^n\right],\nonumber\\
\mathbf{h}_x^{n+0.5}&=\mathbf{\tilde{h}}_x^{n}+\alpha\mathbf{M}_{\mu_x^{-1}}\left[\mathbf{P}_z\mathbf{e}_y^n-\mathbf{P}_y\mathbf{e}_z^{n+0.5}\right]\nonumber\\
\mathbf{h}_y^{n+0.5}&=\mathbf{\tilde{h}}_y^{n}+\alpha\mathbf{M}_{\mu_y^{-1}}\left[\mathbf{P}_z\mathbf{e}_x^n+k_{x,m}\mathbf{e}_z^{n+0.5}\right],
\end{align}
where $\mathbf{\tilde{h}}_x^n, \mathbf{\tilde{h}}_y^n$ are auxiliary vectors and operator matrix $\mathbf{W}_e$ is a sparse one,
\begin{align}\label{Eq51a}
\mathbf{W}_e&=\mathbf{I}+\alpha^2
\left[(k_{x,m})^2
\mathbf{M}_{\epsilon_z^{-1}}\mathbf{M}_{\mu_y^{-1}}+\mathbf{M}_{\epsilon_z^{-1}}\mathbf{P}_y^{*}\mathbf{M}_{\mu_x^{-1}}\mathbf{P}_y
\right].
\end{align}
The field components $\mathbf{e}_x^{n+1},\mathbf{e}_y^{n+1},\mathbf{h}_z^{n+1}$ at time level $n+1$ can be found as
\begin{align}\label{Eq50b}
\mathbf{\tilde{e}}_x^{n+0.5}&=\mathbf{e}_x^{n}+\alpha\mathbf{M}_{\epsilon_x^{-1}}\left[\mathbf{P}_z^{*}\mathbf{h}_y^{n+0.5}-\mathbf{P}_y^{*}\mathbf{h}_z^{n}-\mathbf{j}_x^{n+0.5}\right],\nonumber\\
\mathbf{\tilde{e}}_y^{n+0.5}&=\mathbf{e}_y^{n}+\alpha\mathbf{M}_{\epsilon_y^{-1}}\left[\mathbf{P}_z^{*}\mathbf{h}_x^{n+0.5}-k_{x,m}\mathbf{h}_z^{n}-\mathbf{j}_y^{n+0.5}\right],\nonumber\\
\mathbf{h}_z^{n+1}&=\mathbf{h}_z^{n}+2\alpha\mathbf{W}_h^{-1}\mathbf{M}_{\mu_z^{-1}}\left[-k_{x,m}\mathbf{\tilde{e}}_y^{n+0.5}+\mathbf{P}_y\mathbf{\tilde{e}}_x^{n+0.5}\right],\nonumber\\
\mathbf{e}_x^{n+1}&=\mathbf{\tilde{e}}_x^{n+0.5}+\alpha\mathbf{M}_{\epsilon_x^{-1}}\left[\mathbf{P}_z^{*}\mathbf{h}_y^{n+0.5}-\mathbf{P}_y^{*}\mathbf{h}_z^{n+1}-\mathbf{j}_x^{n+0.5}\right]\nonumber\\
\mathbf{e}_y^{n+1}&=\mathbf{\tilde{e}}_y^{n+0.5}+\alpha\mathbf{M}_{\epsilon_y^{-1}}\left[\mathbf{P}_z^{*}\mathbf{h}_x^{n+0.5}-k_{x,m}\mathbf{h}_z^{n+1}-\mathbf{j}_y^{n+0.5}\right],
\end{align}
where $\mathbf{\tilde{e}}_x^{n+0.5}, \mathbf{\tilde{e}}_y^{n+0.5}$ are auxiliary vectors and operator matrix $\mathbf{W}_h$ is a sparse one,
\begin{align}\label{Eq51b}
\mathbf{W}_h&=\mathbf{I}+\alpha^2
\left[(k_{x,m})^2
\mathbf{M}_{\mu_z^{-1}}\mathbf{M}_{\epsilon_y^{-1}}+\mathbf{M}_{\mu_z^{-1}}\mathbf{P}_y\mathbf{M}_{\epsilon_x^{-1}}\mathbf{P}_y^{*}
\right].
\end{align}

In the staircase approximation of the boundary, the material matrices $\mathbf{M}_{\mu^{-1}}$, $\mathbf{M}_{\epsilon^{-1}}$ are diagonal, and the operators (\ref{Eq51a}),(\ref{Eq51b}) are tri-diagonal matrices. The equations involving these operators can be solved easily by the ``elimination" method~\cite{Sam2001}. However, the staircase approximation of the boundary results in first order convergence in the wake potential. In order to obtain second order convergence and avoid time step reduction, we use conformal method in the same way as described for the rotationally symmetric case in~\cite{Zag03}.
 
In the original problem formulation (\ref{Eq01}), we assumed that the structure walls are perfectly conducting. However, in order to use the Fourier expansion (\ref{Eq04}), it is sufficient to require only that the walls at $x=0$ and $x=2w$ be perfectly conducting. The bottom and top walls of the structure can have different boundary conditions; for example, they can be specified to be metallic with finite resistivity. A similar approach for the case of rotationally symmetric geometry and the use of the staircase approximation for the boundaries was described in~\cite{Tsa12}. In ECHO(2D) we have realized a conformal version of the method for including boundaries with finite conductivity. Details of the implementation of our algorithm will be published elsewhere.
 
\section{Numerical Examples}\label{sec:9} 
 
In this section we consider two example calculations: a corrugated pipe (``dechirper") with perfectly conducting walls, and a tapered collimator with resistive walls. We compare results obtained with ECHO(2D) with those of several 3D codes.

For our dechirper example we used the parameters of the dechirper experiment performed at the Pohang Accelerator Laboratory~\cite{Emma14}. The geometry is shown in Fig.~\ref{Fig01}a. The dechirper structure is 1~m long, has a width $2w=50$~mm, and is made of aluminum. The corrugations are characterized by the period $p=0.5$~mm, height $h=0.6$~mm, and aspect ratio $h/t=2$. The gap between the jaws is adjustable; at a nominal setting the full gap $2a=6$~mm. 

\begin{figure}[htbp]
\centering
\includegraphics*[height=60mm]{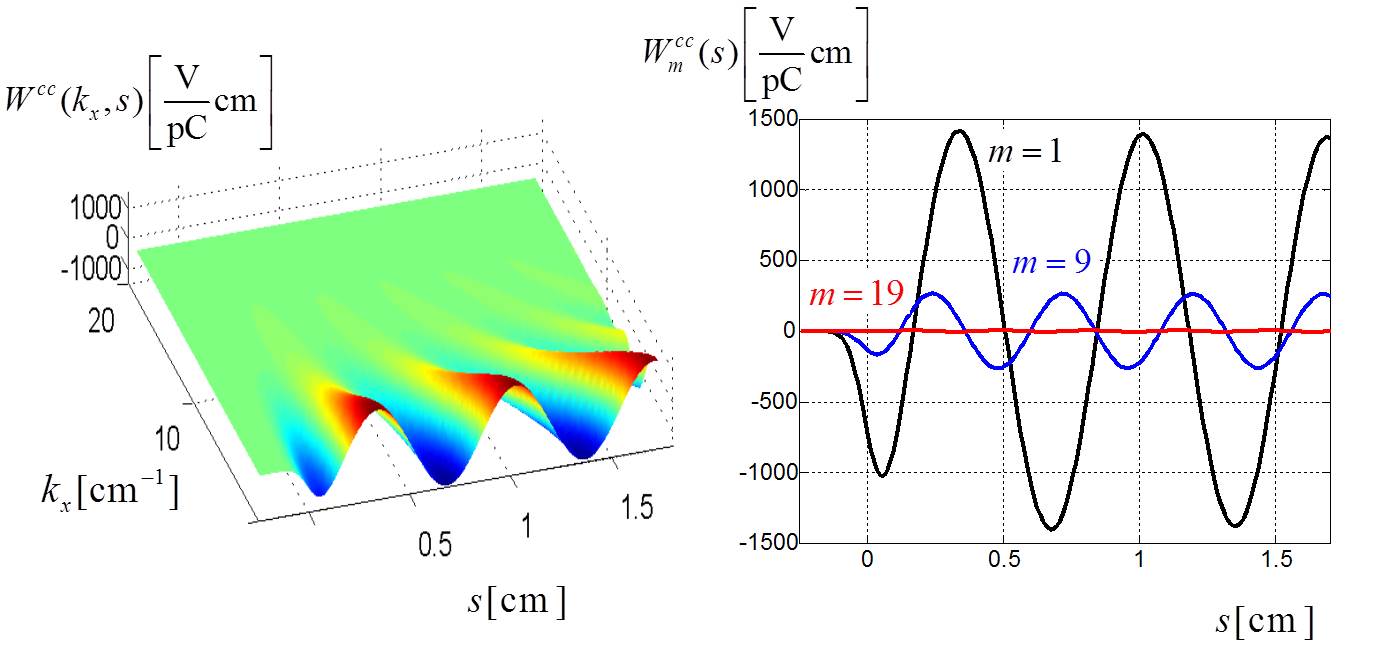}
\caption{For the Pohang dechirper: function $W^{cc}(k_x,s)$ (left), and the harmonics $m=1$, 9, 19, of $W_m^{cc}(s)$ (right).}\label{Fig05}
\end{figure}

\begin{figure}[htbp]
\centering
\includegraphics*[height=60mm]{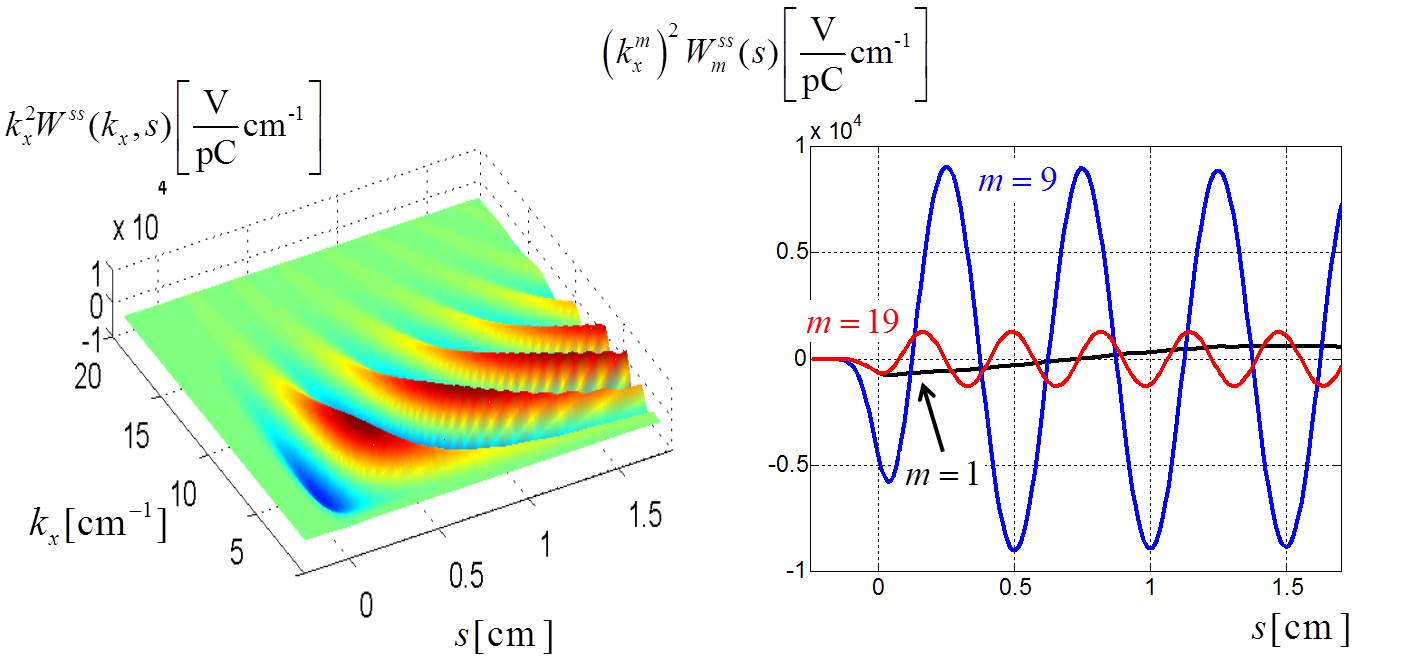}
\caption{For the Pohang dechirper: function $k_x^2W^{ss}(k_x,s)$ (left), and the harmonics $m=1$, 9, 19, of $k_{x,m}^2W_m^{ss}(s)$ (right).}\label{Fig06}
\end{figure}

\begin{figure}[htbp]
\centering
\includegraphics*[height=60mm]{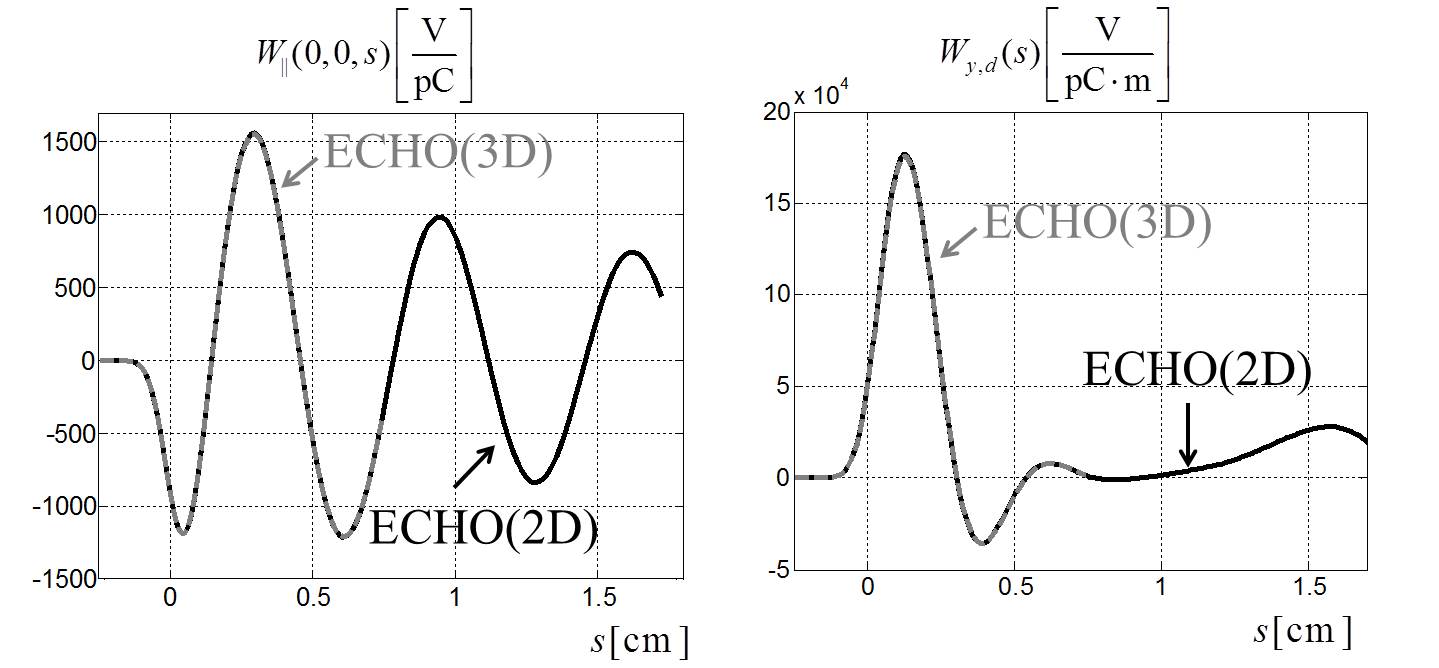}
\caption{Longitudinal wake potential (left) and dipole wake potential (right), calculated from first 15 odd harmonics.}\label{Fig07}
\end{figure}

We have performed calculations of the wakes for a Gaussian bunch with rms length $\sigma_z=0.5$~mm. The transverse dimensions of the bunch were assumed negligibly small. In the right plots of Figs.~\ref{Fig05}--\ref{Fig06}, we show some of the harmonics  $W_m^{cc}(k_x,s)$ and $k_{x,m}^2 W_m^{ss}(k_x,s)$, calculated for $2w=5$~cm. If the bunch passes through the middle of the structure, at $x_0=w$, then only odd harmonics are excited.  In the left plots of  Figs.~\ref{Fig05}--\ref{Fig06}, these functions are shown as continuous functions of $k_x$ (denoted with subscript $m$ removed); they are independent (see Appendix) of the width parameter $2w$. (Here we have used $2w=10$~cm to obtain a denser sampling of the functions in the figures.) Finally, in Fig.~\ref{Fig07}, we compare the longitudinal (left plot) and the dipole (right plot) wakes obtained with ECHO(2D) with the existing, fully 3D code, ECHO(3D)~\cite{Zag05}.  In the calculation of the wake potentials, where the width $2w=5$~cm, we used only the first 15 odd harmonics. We see that the agreement with the existing code is very good. 


In Table~\ref{Table01} we compare the calculated average wakes (the loss factor in the longitudinal case and the kick factors in the transverse cases) with numerical results obtained with the time-domain code T3P of the code suite ACE3P~\cite{Ng13}. The results in the table, denoted by $r$, have been normalized to analytical results, valid for infinitesimally small corrugations (see \cite{Emma14}). We see good agreement except for the case of the dipole wake. (Note, however, that we have good reason to believe that $r$ should always be $\leq1$.) Incidentally, it should be noted that in~\cite{Emma14} the strength of the wakes of the Pohang dechirper were measured, both the longitudinal and transverse wakes, and it was estimated that their measured (effective) $r$ was about 0.75, a result that is in good accord with the ECHO(2D) results.

\begin{table}[htbp]
\centering
\caption{For the Pohang dechirper, for a Gaussian bunch with $\sigma_z=0.5$~mm: the ratio of the numerically obtained and the analytical result(see \cite{Emma14}), that we denote by $r$. We compare the results given in Ref.~\cite{Ng13}, $r$[T3P], with the new results, $r$[ECHO].}
\label{Table01}
\begin{tabular}{lcc}
{\bf Wake}& $r$[T3P]&$r$[ECHO]\\
Longitudinal, loss factor  & 0.84 & 0.83 \\
Dipole, kick factor       & 1.08 & 0.79 \\
Quad, kick factor   & 0.73 & 0.73 \\
\end{tabular}
\end{table}

\begin{figure}[htbp]
\centering
\includegraphics*[height=60mm]{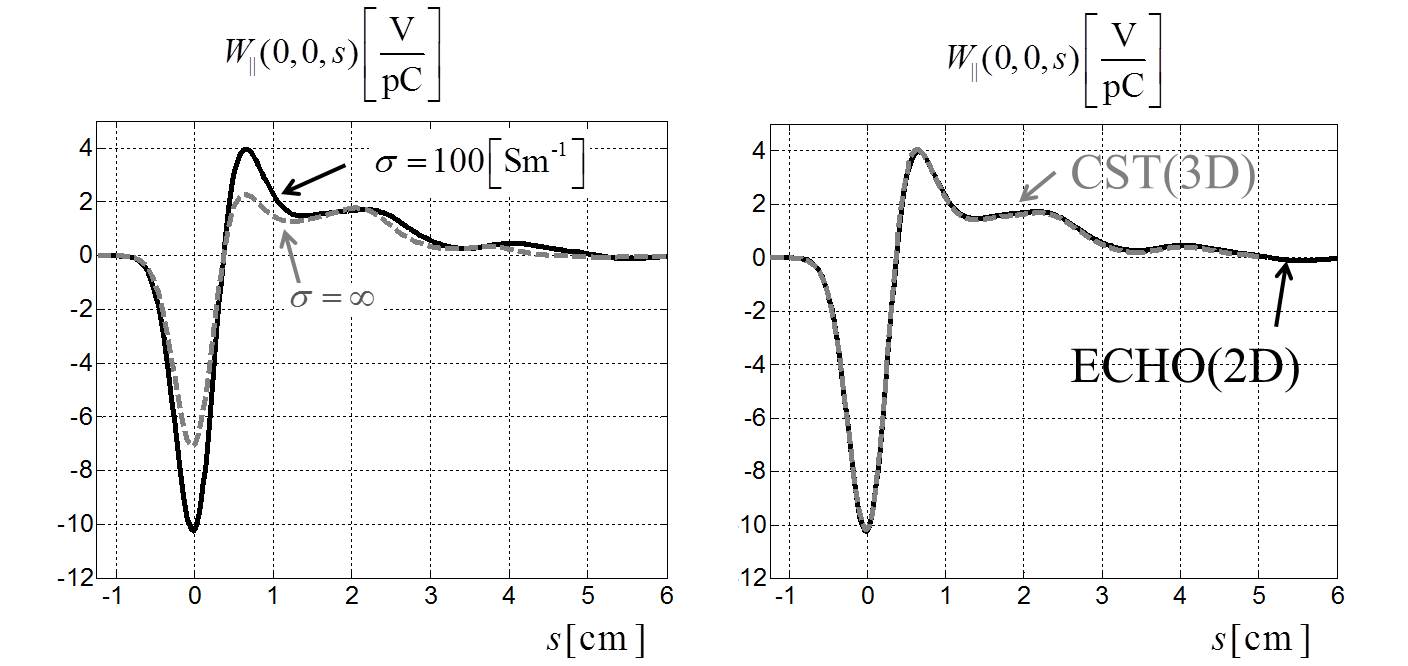}
\caption{Longitudinal wake potential of tapered collimator.}\label{Fig08}
\end{figure}

As the second example problem we consider a symmetric, tapered collimator (see Fig.~\ref{Fig01}b). The dimensions are: width $2w=10$~cm, height of large pipe $2a=10$~cm, length of tapers $T=5$~cm; height  and length of the minimum gap section, $2b=2$~cm and $L=12$~cm. As was mentioned above, ECHO(2D) is capable of modelling structures with metallic walls of finite conductivity. The tapered walls and the walls of the minimum gap section are taken to have conductivity $\sigma=100$~S/m (a poor conductor), while the remaining surfaces are assumed to be perfectly conducting. In the left plot of Fig.~\ref{Fig08} we compare the longitudinal wake for this collimator with one that has the same geometry but is perfectly conducting. The Gaussian bunch in the simulations has an rms length $\sigma_z=0.25$~cm. Both wakes were obtained with ECHO(2D). In the right plot of Fig.~\ref{Fig08} we compare the ECHO(2D) wake potential with the one obtained using a fully three-dimensional, commercially available code CST~\cite{CST}. The good agreement between the results indicates a good  accuracy of the conformal meshing and the resistive wall modelling in ECHO(2D).

\begin{figure}[htbp]
\centering
\includegraphics*[height=60mm]{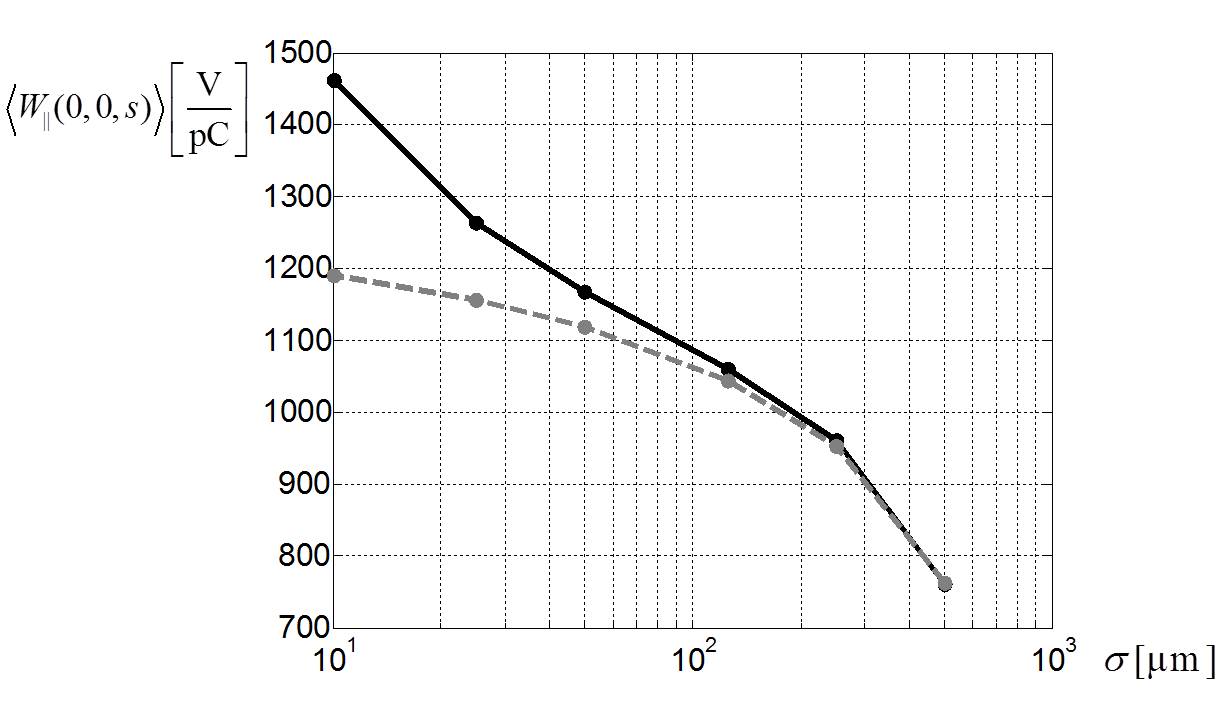}
\caption{Loss factor for very short bunches in the dechirper (in black). The grey symbols give the difference between the loss factors for 2 m-long structure and a 1 m-long structure.}\label{Fig09}
\end{figure}

Finally, in order to demonstrate the capabilities of the new code to calculate wakes of very short bunches, we give in Fig.~\ref{Fig09} the calculated loss factor $\varkappa=\langle W_\parallel(0,0,s)\rangle$ as function of bunch length in the 1~m-long Pohang dechirper (black plotting symbols). Note that the shortest bunch length used in the calculations is $\sigma_z = 10\,\,\mu$m. For the shorter bunches, the loss factor in a 1~m-long structure is larger than the steady-state (periodic) result. We can see that by plotting on the same figure the difference of $\varkappa$ for a 2~m structure minus that for a 1~m structure (the gray curve). Note that the analytical, asymptotic value of the loss factor, for $\sigma_z\rightarrow0$, is $\varkappa=Z_0cL/(2\pi a^2)\cdot(\pi^2/16)=1234.$~V/pC [we have taken $Z_0=377$~$\Omega$, structure length $L=1$~m, and half-gap $a=3$~mm], which agrees well with the linear extrapolation of the ECHO(2D) steady-state results (the gray curve). 

%
\section{Summary}\label{sec:10}
%

In this paper we presented a new method for solving electromagnetic problems and calculating wakefields excited by a relativistic bunch in a structure that is characterized by a rectangular cross-section whose height can vary as function of longitudinal coordinate but whose width and side walls remain fixed. Using the Fourier expansion of the fields, currents and charge densities, we derived  a Fourier representation of the wake potential in terms of four one-dimensional  functions for each harmonic.  We proved that the longitudinal wake is a harmonic function with respect of the coordinates of the leading charge. For a structure that has a horizontal symmetry plane we also proved an important symmetry relation of the wake function.

A numerical method was proposed  for solving Fourier harmonics of the fields. The  method does not generate dispersion in the longitudinal direction and it conserves the energy and charge  in the calculations. The computer resources required to solve the system with several tens of the lowest harmonics are moderate and comparable to those needed for 2D rotationally symmetric calculations. 

%
\section*{Acknowledgements}
%

The authors thank M.Dohlus for helpful discussions. This work was supported by the Department of Energy, contract DE-AC02-76SF00515.

\appendix

\section{Universal continuous functions for structures of constant width}

{
The modal charge distribution, Eq.~(\ref{Eq24}), in Maxwell's equations, 
Eqs.~(\ref{Eq04}), is independent of the structure width, $2w$. The same holds 
for the boundary conditions. Hence the modal solution depends on $2w$ only 
through the harmonic number $k_{x,m} $.  This fact allows us to consider 
components of the electromagnetic fields as continuous functions of the harmonic 
number, $k_x$. The same is true for the longitudinal wake potential. If the 
continuous function  $W_{\parallel}(y_0,k_x,y,s)$, $k_x\in[0,\infty)$, is 
known, then 	we can find the solution for any arbitrary width $2w_1$  as
\begin{equation}\label{Eq06}
	W_{\parallel}(x_0,y_0,x,y,s)=\frac{1}{w_1}\sum\limits_{m=1}^\infty 
W_{\parallel}(y_0,k_{x,m},y,s)\sin(k_{x,m} x_0)\sin(k_{x,m} x),\qquad k_{x,m} 
=\frac{\pi}{2w_1}m.
\end{equation}

In order to find the function  $W_{\parallel}(y_0,k_x,y,s)$, 
$k_x\in[0,\infty)$, we can proceed in the following way. We take width $2w$ 
large enough in order that the sampling of  $W_{\parallel}(y_0,k_x,y,s)$ with 
step  $k_{x,m+1}-k_{x,m} = {\pi}/{(2w)}$ is dense enough for smooth 
interpolation of the function between the discrete values $k_{x,m} = 
{\pi}m/{(2w)}$. Due to the existence of perfectly conducting walls at $x=0$   
and $x=2w$, the discrete Fourier harmonics do not contain a ``zero" harmonic 
with $k_{x,0}=0$. At the same time it is reasonable to find the solution to this 
problem as well. Then, using interpolation, we obtain a universal continuous 
function $W_{\parallel}(y_0,k_x,y,s)$, $k_x\in[0,\infty)$, which is independent 
from the halfwidth $w$.
}


\begin{thebibliography}{1}

\bibitem{Zotter98}
B.W. Zotter, S.A. Kheifets,  {\it Impedances and Wakes in High-Energy Particle Accelerators} (World Scientific, London, 1998).

\bibitem{Pukhov99}
A. Pukhov, J. Plasma Physics {\bf 61}, 425 (1999).

\bibitem{Novo05}
A. N. Novokhatski, SLAC Report No. SLAC-PUB-11556, 2005.

\bibitem{Zag05}
I.A. Zagorodnov, T. Weiland, Phys. Rev.  ST Accel. Beams {\bf 8}, 042001 (2005).

\bibitem{Bane12}
K.L.F. Bane and G. Stupakov, Nucl. Instrume. and Methods {\bf 690}, 106 (2012).

\bibitem{Paech07}
A.R. Paech, {\it Evaluation numerischer Methoden zur Berechnung von Synchrotron-strahlung am ersten Bunchkompressor des Freie-Elektronen-Lasers FLASH} (Dissertation, TU Darmstadt, 2007).

\bibitem{Wei92}
T. Weiland, R. Wanzenberg, Lect. Notes Phys. {\bf 400}, 39 (1992). 

\bibitem{Zag06}
I. Zagorodnov, Phys. Rev.  ST Accel. Beams {\bf 9}, 102002 (2006).

\bibitem{Zag02}
T. Weiland, I. Zagorodnov, J. Comput. Phys. {\bf 180}, 297 (2002).

\bibitem{Gluckstern}
R. Gluckstern, J. Van Zeijts, B. Zotter, Phys. Rev. E {\bf 47}, No.
1, 656 (1993).

\bibitem{Bane07}
K.L.F. Bane, G. Stupakov, I. Zagorodnov, Phys. Rev.  ST Accel. Beams {\bf 10},  074401 (2007).

\bibitem{Wei96}
T. Weiland, Int. J. Numer. Model. {\bf 9}, 295 (1996).

\bibitem{Yee66}
K.S. Yee, IEEE Trans. Antennas and Propagation {\bf 14}, 302 (1966).

\bibitem{Thoma97}
P. Thoma, {\it Zur numerischen Loesung der Maxwellschen Gleichungen im Zeitbereich} (Dissertation, TU Darmstadt, 1997).

\bibitem{Sam2001}
A.A. Samarskii, {\it The Theory of Difference Schemes} (Marcel Dekker, Inc., New York, 2001).

\bibitem{Zag03}
I. Zagorodnov, R. Schuhmann, T. Weiland, J. Comput. Phys. {\bf 191}, 525 (2003).

\bibitem{Tsa12}
A. Tsakanian, M. Dohlus, I. Zagorodnov, Phys. Rev. ST Accel. Beams {\bf 15}, 054401 (2012).

\bibitem{Emma14}
P. Emma et al, Phys, Rev. Letters {\bf 112},  034801 (2014).

\bibitem{Ng13}
C.-K. Ng, K.L.F. Bane, in Proceedings of PAC 2013, Pasadena, CA USA,WEPAC46, 2013. 

\bibitem{CST}
CST-Computer Simulation Technology, CST Particle Studio, http://www.cst.com.

\end{thebibliography}
\end{document}